\documentclass[manuscript,nonacm]{acmart}
\AtBeginDocument{%
  }

\copyrightyear{2026}
\acmYear{2026}
\setcopyright{cc}
\setcctype{by}
\acmConference[CHI '26]{Proceedings of the 2026 CHI Conference on Human Factors in Computing Systems}{April 13--17, 2026}{Barcelona, Spain}
\acmBooktitle{Proceedings of the 2026 CHI Conference on Human Factors in Computing Systems (CHI '26), April 13--17, 2026, Barcelona, Spain}
\acmPrice{}
\acmDOI{10.1145/3772318.3791800}
\acmISBN{979-8-4007-2278-3/2026/04}

\begin{document}

\title{\revision{Flow on Social Media? Rarer Than You'd Think}}

\author{Michael T. Knierim}
\affiliation{%
  \institution{Karlsruhe Institute of Technology}
  \country{Germany}
}
\affiliation{%
  \institution{University of Nottingham}
  \country{UK}
}

\author{Thimo Schulz}
\affiliation{%
  \institution{Karlsruhe Institute of Technology}
  \country{Germany}
}

\author{Moritz Schiller}
\affiliation{%
  \institution{Karlsruhe Institute of Technology}
  \country{Germany}
}

\author{Jwan Shaban}
\affiliation{%
  \institution{University of Nottingham}
  \country{UK}
}

\author{Mario Nadj}
\affiliation{%
  \institution{University of Duisburg-Essen}
  \country{Germany}
}

\author{Max L. Wilson}
\affiliation{%
  \institution{University of Nottingham}
  \country{UK}
}

\author{Alexander Maedche}
\affiliation{%
  \institution{Karlsruhe Institute of Technology}
  \country{Germany}
}

\renewcommand{\shortauthors}{Knierim et al.}

\newcommand{\revision}[1]{\textcolor{black}{#1}}
\newcommand{\revisionfinal}[1]{\textcolor{black}{#1}}

\begin{abstract}
Researchers often attribute social media’s appeal to its ability to elicit flow experiences of deep absorption and effortless engagement. Yet prolonged use has also been linked to distraction, fatigue, and lower mood. This paradox remains poorly understood, in part because prior studies rely on habitual or one-shot reports that ask participants to directly attribute flow to social media. To address this gap, we conducted a five-day field study with 40 participants, combining objective smartphone app tracking with daily reconstructions of flow-inducing activities. Across 673 reported flow occurrences, participants rarely associated flow with social media (2\%). Instead, heavier social media use predicted fewer daily flow occurrences. We further examine this relationship through the effects of social media use on fatigue, mood, and motivation. Altogether, our findings suggest that flow and social media may not align as closely as assumed - and might even compete - underscoring the need for further research.
\end{abstract}



\maketitle

\section{Introduction}
Flow - the state of deep absorption and optimal engagement - is often linked to better mood, well-being, and performance in daily life \cite{Csikszentmihalyi1975, Abuhamdeh2020, Peifer2022}. In HCI, flow functions both as an analytic lens and a design aspiration: systems that help people focus, progress, and feel effective are widely valued.
As one of the most pervasive digital systems in everyday life, social media have attracted significant interest as potential sources of flow-like engagement \cite{Cipresso2015, Mauri2011, Zhao2023, Liu2024}. The interactive nature of these systems - with their constant streams of content, immediate feedback through likes and comments, and opportunities for creative expression - appears to offer many of the conditions that flow theory suggests are necessary for optimal experience \cite{Cipresso2015, Mauri2011}. Conversely, studies have documented that people report absorption, focused attention, and enjoyment while using platforms like Facebook, Instagram, and TikTok \cite{Zhao2023, Liu2024} - and some research has even suggested that flow experience in social media might be fueling addictive tendencies \cite{Brailovskaia2020, Chen2025}.

However, research also finds that extensive social media use can lead to reduced mood, greater distraction, and heightened fatigue \cite{Kushlev2019, Verduyn2017, Beyens2024, Valkenburg2022, Ernala2022} - experiences that stand in contrast with typical flow-preconditions and consequences. This creates a fundamental tension: if social media regularly facilitates the kinds of positive, absorptive experiences that characterize flow, why do we observe these adverse effects? Moreover, the design features that platforms use to maintain engagement - infinite scroll, variable reward schedules, rapid context switching - seem at odds with the active engagement behavior that traditional flow theory emphasizes \cite{Engeser2016, Csikszentmihalyi1989}.
This paradox remains poorly understood, in part because of methodological challenges in how flow has been studied in social media contexts. Much prior work operationalizes flow through individual facets like absorption or time distortion, rather than capturing the full theoretical construct with its requirements for challenge–skill balance, clear goals, and unambiguous feedback \cite{Csikszentmihalyi1975, Engeser2008, Abuhamdeh2020}. In addition, many studies rely on self-reports that explicitly prompt participants to consider whether they experienced flow during social media use with continuous scales, or employ platform-specific flow scales that presuppose this connection \cite{Kaur2016b, Mauri2011, Cipresso2015}. Such approaches risk overstating the link between flow and social media, as participants are effectively required to map their flow responses onto platform use - even if this does not reflect how flow naturally arises in their daily lives.
These methodological concerns point to three fundamental research questions: 

\begin{itemize}
    \item \textbf{RQ1 - Prevalence:} When people independently reflect on their day, how often do they \emph{spontaneously} identify social media use as flow-inducing?
    \item \textbf{RQ2 - Contribution:} Does social media use meaningfully add to people’s daily tally of flow?
    \item \textbf{RQ3 - Pathways:} Which psychological states (e.g., readiness, fatigue) accompany or explain the link between social media use and flow?
\end{itemize}

To answer these questions, we conducted a five-day field study designed to decouple experience from attribution. We paired app-level telemetry of social media use with independent daily reconstructions of flow-inducing activities using an adapted Day Reconstruction Method (DRM) \cite{Kahneman2004}. Participants freely reported flow experiences across the day without being primed to consider social media (but encouraged to consider all activities in work and leisure), while app-usage logs captured their actual social media activity. This combination - similar to recent work that links platform logs with self-reports to disentangle objective behavior from subjective experience \cite{Ernala2022} - allowed us to examine both within-person dynamics and between-person differences in how social media use relates to flow.
By testing rather than presuming the link between social media and flow, our study provides methodological and empirical contributions - initial evidence on how rarely social media is experienced as flow-inducing and how usage patterns relate to everyday optimal experience. Further replications and methodological extensions will be valuable, for instance by integrating cross-device logging, combining DRM with experience sampling or physiological measures, and examining within-app behaviors and contexts. So far, our findings already illustrate the potential of pairing telemetry with independent flow reports and help refine our understanding of how social media platforms relate to positive psychological experiences such as flow.

\section{Related Work}
\subsection{Flow in Everyday Life}
Flow is a state of deep, effortless absorption in a task, typically accompanied by fluent action and linked to enhanced performance and well-being \cite{Csikszentmihalyi1975, Bruya2010, Oliveira2019, Yotsidi2018, Afergan2014, Schick2025}.
%
Classic accounts hold that flow is most likely when three conditions align: the task's challenge matches the individual's skills, goals are clear, and feedback is unambiguous \cite{Csikszentmihalyi1975}. Because many activities satisfy these conditions, flow has been studied across work settings \cite{Brown2023, Afergan2014, Knierim2025, Knierim2019}, education \cite{Wang2023, Pastushenko2020}, games \cite{Klarkowski2016, Labonte2016, Klarkowski2015, Tozman2015, Johnson2015}, and interactions with digital technologies \cite{Bian2016, Berger2023}.
Notably, people report flow more often at work than during passive leisure - such as watching TV - despite the latter's hedonic appeal, a pattern often referred to as the "paradox of work" \cite{Engeser2016, Quinn2005, Csikszentmihalyi1989}. In extension, researchers have also highlighted that flow experiences at work are related to the quality of recovery in the previous evening \cite{Debus2014} and that flow appears to be linked to mental fatigue and readiness \cite{Schick2025}.
%
Within HCI, as it is associated with great user experiences, flow has been prolifically studied for example in single-player \cite{Klarkowski2015, Tozman2015, Ewing2016} and multi-player gaming \cite{Labonte2016, Johnson2015}, virtual reality experiences \cite{Bian2016}, live-streaming \cite{Wang2022}, music-streaming services \cite{Wang2014}, educational systems \cite{Oliveira2019}, and adaptive work software \cite{Afergan2014}.
The study of flow in social media contexts has emerged as a significant area of HCI-research, though with varied approaches and conclusions. 

\subsection{Flow in Social Media Use}
A substantial body of work frames social media platforms as conducive to flow-like engagement, drawing on flow theory to explain user behavior and platform effects. Yet this literature spans several distinct traditions that differ in assumptions, measures, and findings.
%
One strand invokes flow primarily as an explanatory framework for attention capture and platform “stickiness,” often without directly measuring momentary states. For instance, studies on the academic consequences of Facebook use or analyses of attention-capture designs draw on flow to explain why users remain absorbed, but focus on downstream outcomes rather than the flow experience itself \cite{Wang2018,Monge2023}. This approach treats flow as a plausible mechanism underlying observed behavioral patterns - such as extended usage sessions or difficulty disengaging.
A second research tradition directly measures the relationship between social media use and flow. Psychophysiological studies document absorption-like patterns during Facebook use, including reduced cortisol levels and activation patterns consistent with focused engagement \cite{Mauri2011,Cipresso2015}. Survey research finds that flow-prone individuals are more susceptible to problematic social media use, with flow mediating between positive platform attitudes and addictive behaviors \cite{Brailovskaia2020,Wickord2025}. These findings suggest that flow experiences can both enrich and potentially contribute to excessive platform engagement.
A third line of research advances theory-driven accounts that position flow as central to platform design and user experience. Work on short-video platforms, for example, proposes that flow states sustain viewing and reduce self-control, sometimes via willpower depletion \cite{Kaur2016a,Kaur2016b,Yang2014,Zhou2025,Chen2025}. These studies stress how design features such as personalized feeds and seamless transitions may foster absorption, encouraging continued use.

However, despite frequent claims that social media supports flow, it remains unclear whether the platform experiences people report actually constitute authentic flow.
Some platform features can clearly support flow’s preconditions: goal-directed activities like composing posts or editing videos provide challenge–skill balance, immediate feedback through reactions, and clear proximal goals \cite{Mauri2011,Cipresso2015,Kaur2016b}. Consistent with this, interactive exchanges are typically associated with more favorable experiences than passive scrolling \cite{Burke2011,Verduyn2017,Godard2024}.
However, predominant usage patterns can be argued to systematically undermine flow conditions. Passive feed browsing - the modal form of engagement - often lacks clear goals and meaningful challenge, resembling TV viewing in its low flow potential \cite{Csikszentmihalyi1989,Engeser2016}. \revision{Such passive use can even cause dissociative experiences where users scarcely register the browsed content \cite{Baughan2022}.} 
Algorithmic content curation can maintain engagement without fostering concentration \cite{Monge2023}, while notifications, infinite scroll, and rapid context switching fragment attention and may interrupt flow \cite{Peifer2019}. Moreover, social comparison processes can trigger negative affect \cite{Verduyn2015,Verduyn2017, Ernala2022}, and experimental evidence shows that short social media sessions can increase mental fatigue and impair cognitive performance \cite{Gantois2021,Jacquet2023} - states antithetical to flow \cite{Debus2014, Schick2025}.
These features suggest that engaging social media use may often represent a "pseudo-flow" - behaviorally similar through time distortion and absorption, but lacking challenge-skill balance and active engagement that characterize authentic flow.

This tension suggests two pathways through which social media may relate to flow: (i) active, goal-directed, reciprocal use that provides challenge, feedback, and control (supporting flow), and (ii) passive, unstructured scrolling that induces absorption without challenge and is linked to fatigue or diminished affect (hindering true flow). Clarifying which pathway dominates requires methods that can disentangle these mechanisms.
Yet most prior work relies on attributional self-reports or platform-specific flow scales that presuppose a link to social media \cite{Kaur2016b, Cipresso2015, Mauri2011}, making it unclear whether genuine flow occurs. Few studies integrate subjective reports with device telemetry (e.g. app-use tracking) or within-person models that can explain momentary and dynamic behaviors. Moreover, boundary conditions - active vs.\ passive use, content, and goals - further shape whether flow aligns with immersion and enjoyment or instead with distraction, fatigue, and compulsive use \cite{Leung2020,Wickord2025, Whiting2013, Brailovskaia2020}.
Altogether, this highlights the need for approaches that move beyond attribution and capture when and under what conditions flow genuinely co-occurs with daily social media use.

\section{Method}
\subsection{Study Design}
\begin{figure*}[h]
  \centering
  \includegraphics[width=0.85\linewidth]{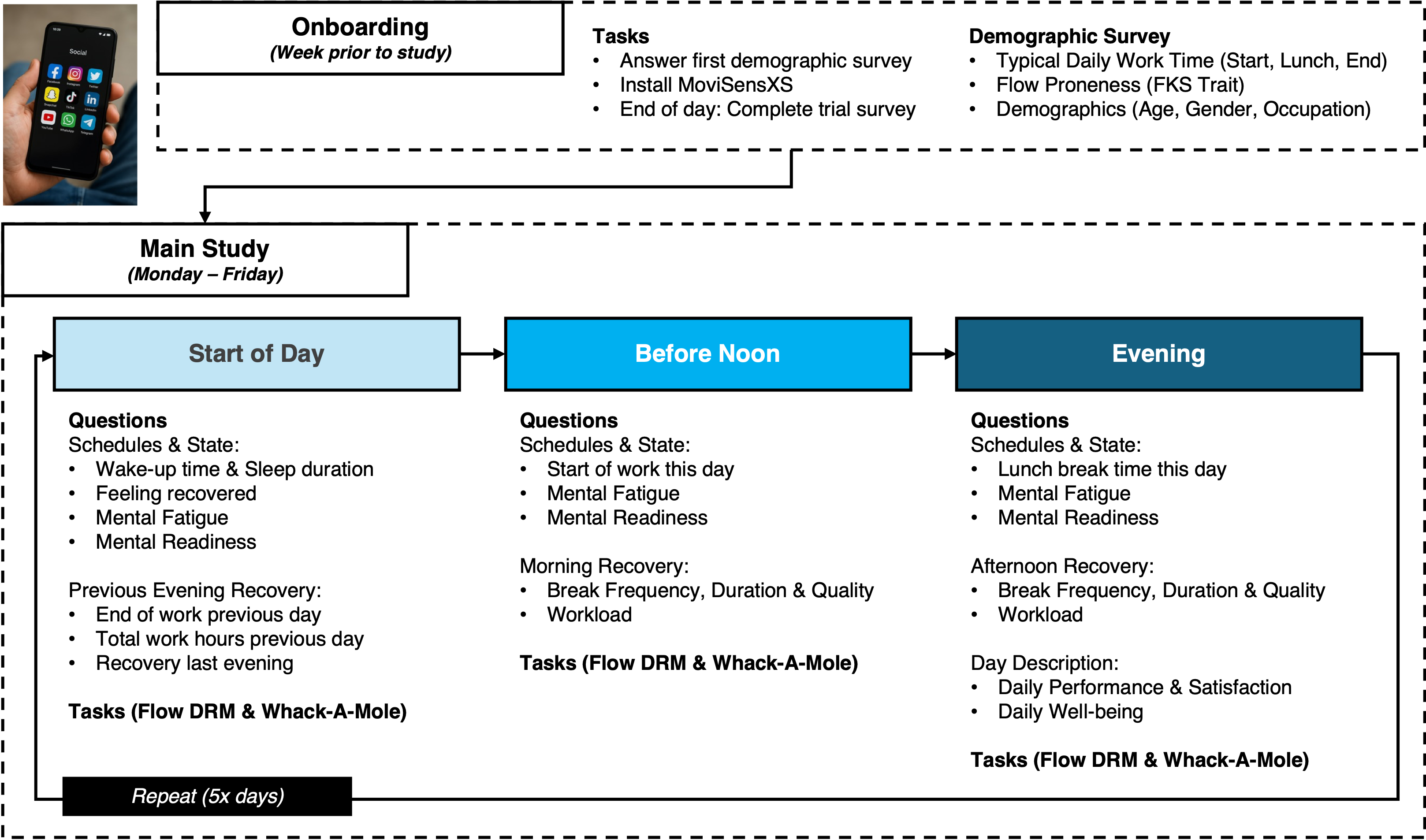}
  \caption{Overview of the study design: participants completed an onboarding session followed by three daily reflection surveys (morning, before noon, evening) across five study days.}
  \label{fig:studydesign}
  \Description{The figure shows the study timeline. After an onboarding session where participants installed the app, the main study ran Monday to Friday with three daily reflection surveys. In the morning, participants reported sleep, recovery, and mental states; before noon, they reported breaks, workload, fatigue, and readiness; in the evening, they reported breaks, daily recovery, fatigue, readiness, performance, satisfaction, and well-being. At each session, they also reported flow activities and completed a short reaction-time task.}
\end{figure*}

We conducted a five-day longitudinal study combining continuous smartphone app tracking with structured reflections. Participants completed three reflection sessions per day (morning, midday, end-of-workday), totaling 15 sessions (see Figure \ref{fig:studydesign}), while their smartphone app use - particularly social media - was logged continuously (details below).

During each reflection session, participants completed surveys on their flow experiences in the preceding hours using an adapted Day Reconstruction Method (DRM)~\cite{Kahneman2004}. The DRM asks participants to systematically reconstruct their activities, their timing, and associated subjective states. This approach addresses limitations of traditional experience sampling, which captures in-the-moment experiences through random notifications~\cite{Trull2013,Csikszentmihalyi2003, Debus2014} but may miss flow episodes occurring between prompts, leading to biased estimates of flow-eliciting activities~\cite{Bartholomeyczik2024}. By contrast, the DRM provides more comprehensive coverage by reconstructing entire time periods rather than isolated moments.  
In this study, we adapted the DRM to cover three temporal windows - the previous evening, the morning, and the afternoon. Segmenting the day in this way reduced cognitive load and supported more accurate recall by breaking reconstructions into manageable blocks. Each session also included brief reaction time tasks, collected for potential future analyses of mental fatigue, but not examined in the present article.

\subsection{Measures}
\subsubsection{Flow Identification via Day Reconstruction Method} 
\begin{figure*}[h]
  \centering
  \includegraphics[width=\linewidth]{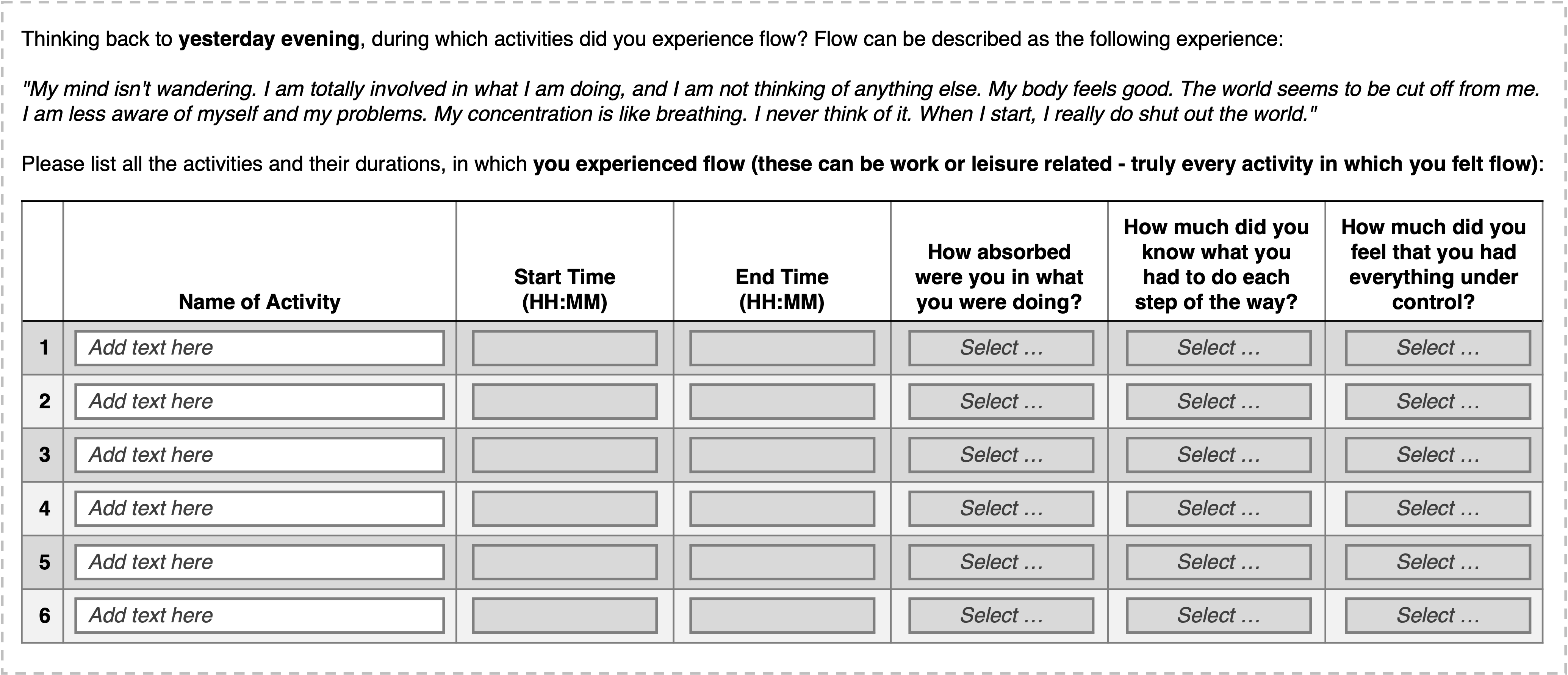}
  \caption{Screenshot of the adapted Day Reconstruction Method (DRM) survey used to capture flow activities.}
  \label{fig:flowDRM}
  \Description{The figure shows an online survey form asking participants to recall activities from the previous evening during which they experienced flow. At the top, a short description of flow is provided as guidance. Below, a table allows participants to enter up to six activities with their start and end times. For each activity, participants rate three aspects on 7-point scales: how absorbed they were, how clearly they knew what to do, and how much they felt in control.}
\end{figure*}

A screenshot of the DRM survey is shown in Figure~\ref{fig:flowDRM}. To orient participants, we included a short descriptive quote illustrating what flow feels like, following established practice in flow research~\cite{Wilson2016}. In each reflection session, participants could report up to six flow-eliciting activities, with fields kept optional to avoid forcing reports when no flow occurred.  
For every reported activity, participants specified its timing and duration and rated the intensity of flow on 7-point Likert scales using the three-item Flow Short Scale~\cite{Bartholomeyczik2024,Engeser2008}.

\subsubsection{Additional Surveys.} 
Beyond flow reports, we collected additional contextual measures (details in Figure~\ref{fig:studydesign}) across the three daily reflection sessions, focusing on mental fatigue, mental readiness, and end-of-day performance, satisfaction, and well-being. Mental fatigue (4-item MFI; \cite{Smets1995}) and mental readiness (composite of 3 items on mood, sleepiness, and motivation; \cite{Ng2023}) were assessed in the morning, midday, and evening. At the end of the day, participants additionally rated their total daily performance and satisfaction (2 items from the NASA-TLX; \cite{Hart1988}) and well-being (using the PERMA Profiler’s five highest-loading items per dimension; \cite{Butler2016}).

\subsubsection{Smartphone App Tracking}
We used MoviSensXS\footnote{https://www.movisens.com/en/products/movisensXS/}, a widely adopted platform for experience sampling, to track participants’ smartphone app usage. In line with GDPR requirements and technical limitations, apps had to be specified a priori rather than logged exhaustively.  
To define this tracking list, we followed a two-stage process. First, 27 students from our target population (9 female, 17 male, age 20–41, median 22) shared screenshots of their most-used apps from the previous week, yielding 48 frequently used applications. Second, two researchers systematically reviewed the Google Play Store to identify additional high-download apps in the same categories, adding eight (Telegram, Threads, Bluesky, Tellonym, Firefox, Tumblr, Threema, Mastodon) that did not appear in the initial sample.

The resulting list of 56 apps was configured for tracking on study participants' smartphones. For analysis purposes we operationalized active vs.\ passive social media use at the app-category level: messaging-dominant apps (e.g., WhatsApp, WeChat) as \emph{active} - a proxy for directed communication - and feed/video-dominant apps (e.g., Instagram, Facebook, TikTok, YouTube) as \emph{passive} - a proxy for content consumption. This follows prior distinctions between directed communication and passive consumption \cite{Burke2011,Verduyn2017,Godard2024} and is in line with related work that has, for pragmatic reasons, grouped apps by primary affordances when only app-level telemetry is available \cite{Schoedel2022,Yu2020,Liang2024}. Because users can engage in both active and passive behaviors within the same app, our categorization should be viewed as a proxy rather than a behavior-level measure.

\subsection{Procedure}
The study was approved by the first author's university’s Institutional Review Board. One week before data collection, participants attended a lab onboarding session, provided informed consent, and installed the MoviSensXS app for tracking and communication. They reported typical sleep and work schedules \cite{Debus2014}, which were used to configure personalized reminder notifications at three daily timepoints: before work, before lunch, and at the end of the workday. Participants also completed demographic questions (age, sex, occupation). To ensure smooth procedures, each participant received a test reminder and completed a practice reflection session the same evening, allowing technical issues to be resolved before formal data collection began the following Monday.

\subsection{Participants}
We recruited 44 university students (10 women, 34 men; age 18–34, median 24, \revision{various nationalities}) from a public student pool \revision{in Germany} during the academic term. \revision{Previous flow experience sampling work \cite{Debus2014, Bartholomeyczik2024, Brown2023} employed samples from $\sim$20-120 participants. We chose our sample size to balance participant burden with data depth (longer reflection through the DRM compared to ESM reports), prioritizing dense within-person sampling over larger between-person N, mainly to enable exploration of the association patterns that our study targets.}
All participants confirmed a standard Monday–Friday daytime schedule (lectures/study/other daytime commitments) and no shift-based employment (e.g., night or rotating shifts) during the study week.
Eligibility required active social media use (>30 minutes/day), no professional social media responsibilities (e.g., influencer, social media manager), regular Android use (for MoviSensXS), and good health without neurological or sleep disorders. Participants received 100 Euro contingent on completing at least 80\% of the scheduled reflection surveys.
Four participants fell below this threshold and were excluded, yielding a final sample of 40. Compliance among retained participants was high: 33 completed all 15 surveys, six completed 14, and one completed 13 (82.5\% full compliance).

\section{Results}
\subsection{Flow Experience Patterns}
\subsubsection{Flow Activities}
Using the DRM, participants could report flow activities for three periods per day (morning, afternoon, evening). We recorded 673 total responses. The median number of reports per participant was 15.5 (min = 3, max = 55); per day, the median per participant was 3 (min = 1, max = 11).

\revisionfinal{To examine reasons for flow, we applied an inductive coding approach - informed by conventional qualitative content analysis \cite{Hsieh2005}. 
We chose such an inductive approach because prior flow studies often rely on broad and context-specific activity categorizations (e.g., work vs. leisure or active vs. passive leisure; \cite{Csikszentmihalyi1989,Engeser2016, Quinn2005}), or were developed in earlier phases of digital media use \cite{Gruner2016}. As our study targets contemporary everyday media experiences, we anticipated that recent and more fine-grained media activity forms may emerge that might not be adequately captured by existing activity schemes.}
Half of the responses were randomly sampled and independently coded by two researchers. After removing duplicates, the researchers 
\revision{met to reconcile divergent codes by refining code definitions and reviewing exemplars from the data until a coherent, jointly agreed-upon scheme was reached;} the remaining responses were then coded by one researcher.

\begin{figure*}[h]
  \centering
  \includegraphics[width=1\linewidth]{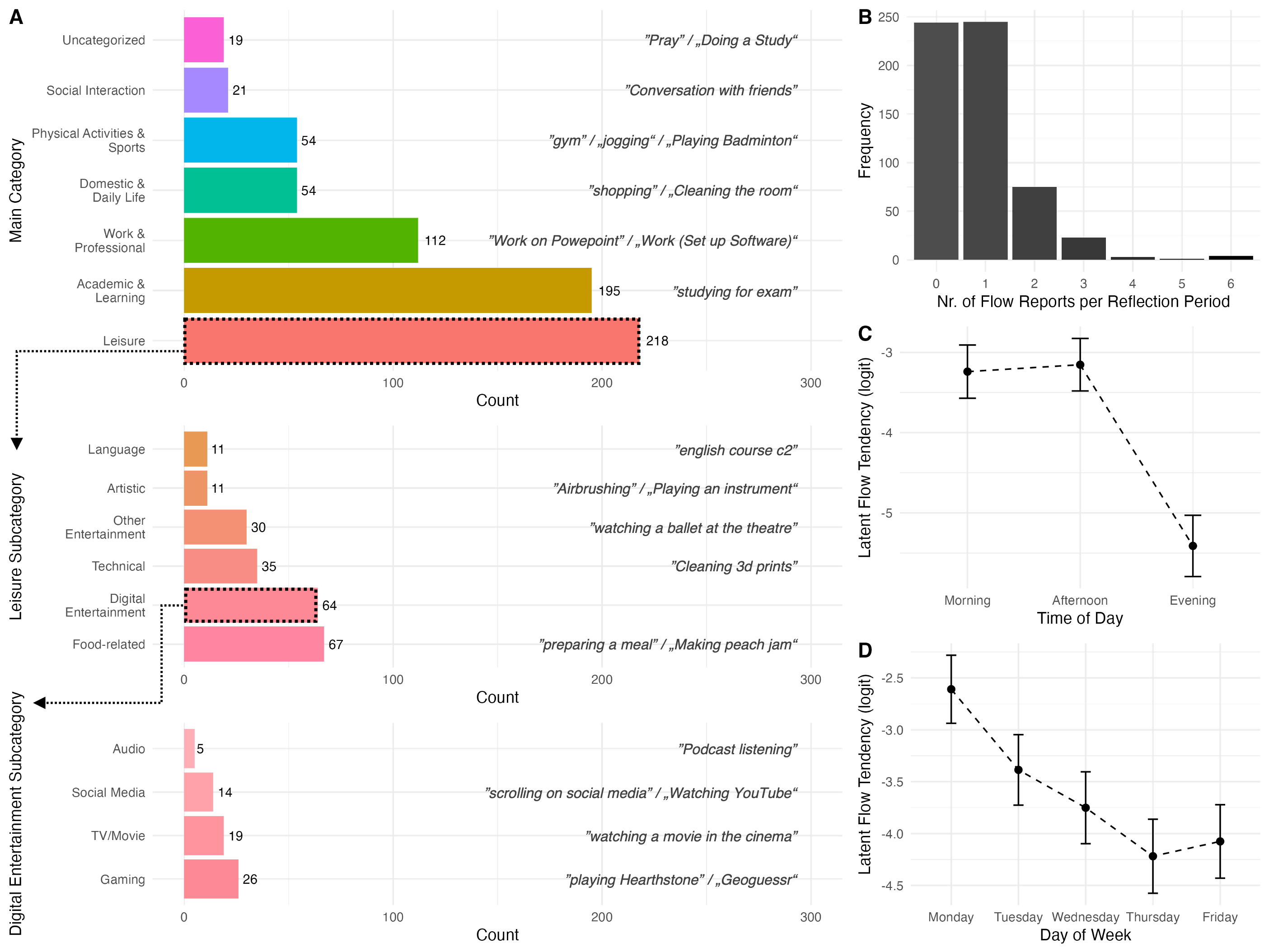}
  \caption{Overview of flow reports by activity categories (left), and their distribution across time of day and week (right). Error bars indicate one SE.}
  \label{fig:flowByCategory}
  \Description{The figure shows several bar charts and line plots summarizing reported flow activities. On the left, stacked bar charts display the number of flow reports by category: most were in Leisure and Academic \& Learning, followed by Work \& Professional. Within Leisure, the largest subcategories were Food-related and Digital Entertainment (including Gaming and TV/Movies), while Social Media was a very small fraction. On the right, histograms show that most participants reported one or two flow episodes per reflection session, most often in the morning and afternoon rather than in the evening. A line plot indicates that flow reports declined gradually from Monday through Friday.}
\end{figure*}

Seven categories emerged: \emph{Leisure} (218), \emph{Academic \& Learning} (195), \emph{Work \& Professional} (112), \emph{Domestic \& Daily Life} (54), \emph{Physical Activities \& Sports} (54), \emph{Social Interaction} (21), and \emph{Uncategorized} (19). 
Within \emph{Leisure}, we derived subcategories for finer-grained analysis: \emph{Food-Related} (67), \emph{Digital Entertainment} (64), \emph{Technical} (35), \emph{Other Entertainment} (30), \emph{Artistic} (11), and \emph{Language} (11). 
All social-media-related flow mentions (14 reports by 8 participants) appeared in \emph{Digital Entertainment}, which also included \emph{Gaming} (26), \emph{TV/Movie} (19), and \emph{Audio} (5) \revision{- all of which reflected consumption (listening to music, audiobooks or podcasts).}
\revision{Social-media related flow reports include watching videos (6 - e.g., “watching YouTube”), feed scrolling (6 - e.g., "scrolling through Instagram"), and unspecified activities (2 - e.g., "Social Media").} See Figure~\ref{fig:flowByCategory}A for categories with example quotes.
Thus, overall, reports of flow during social media use were rare: 14 of 673 reports (2.1\%). This suggests that while flow can occur on social media and is noticed, participants rarely identified such episodes as flow-eliciting in their day.

\subsubsection{Daily \& Weekly Development}
To examine the temporal dynamics of flow, we analyzed its progression across day and week.  
As is common in DRM data~\cite{Kahneman2004, Hedeker2010, Green2021} when tallying rare events, the distribution of flow reports (see Figure \ref{fig:flowByCategory}B) was bounded with many zeros and strong right-skew, suggesting overdispersion and non-Gaussian behavior. 
Because the values can be read as either graded categories or event counts, we fitted both cumulative-logit ordinal mixed models and negative-binomial mixed models. In each case, participants were added as a random effect to account for inter-individual differences and the repeated measurement (within-person dependence).
We report the ordinal results as primary, as the small, bounded scale likely reflects graded categories rather than precise counts; the NB models yielded the same direction and broadly similar effect sizes, supporting robustness.
In summary, regression analyses show a significant decrease of flow in the evening of the day (\(b=-0.941\), \(SE=0.109\), \(p<.001\) - see Figure \ref{fig:flowByCategory}C), and a reduction of flow occurrences throughout the week (\(b=-0.379\), \(SE=0.059\), \(p<.001\) - see Figure \ref{fig:flowByCategory}D).

\subsubsection{Daily Flow Correlates}
\begin{figure*}[h]
  \centering
  \includegraphics[width=1\linewidth]{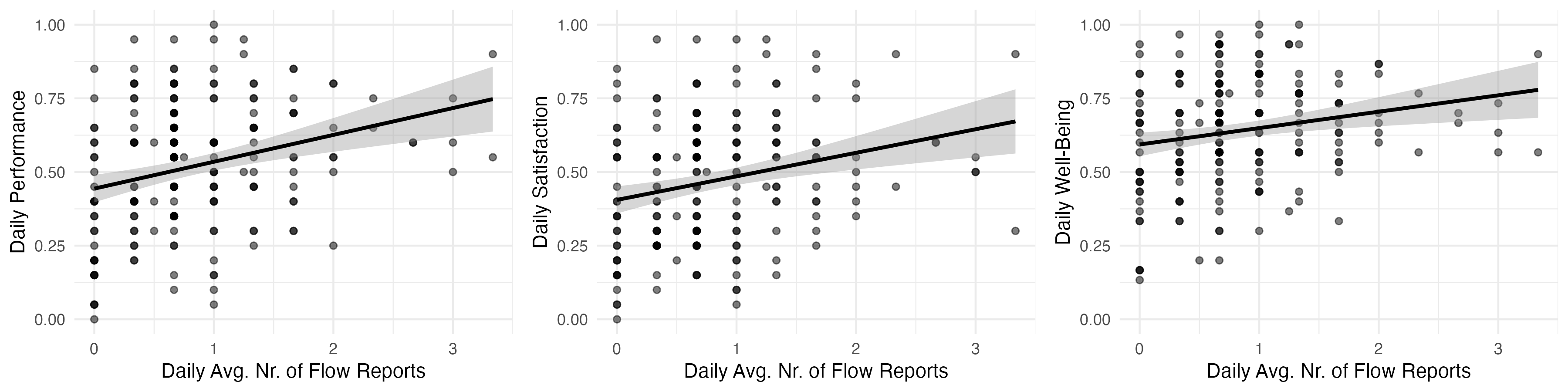}
  \caption{Positive correlations between daily flow reports and end-of-day outcomes: (left) performance, (middle) satisfaction, and (right) well-being.}
  \label{fig:flowDailyExperienceCorrs}
  \Description{The figure shows three scatterplots with regression lines, illustrating the positive relationships between daily flow reports and end-of-day outcomes. In the first panel, more flow reports are associated with higher daily performance; in the second, with greater satisfaction; and in the third, with higher well-being. All regression lines slope upward, with shaded confidence intervals indicating significant positive associations, suggesting that days with more reported flow experiences also tended to be rated more positively on these outcomes.}
\end{figure*}

To test the well-established link between flow and daily outcomes, we aggregated flow reports to daily averages per participant and computed repeated-measures correlations~\cite{Bakdash2017} with evening reports of performance, satisfaction, and well-being. Aggregation improved normality of the variables (Figure~\ref{fig:flowDailyExperienceCorrs}).  
All three outcomes showed significant positive associations with daily flow: performance (\(r(159)=0.26\), \(p<.001\)), satisfaction (\(r(159)=0.18\), \(p=.019\)), and well-being (\(r(159)=0.20\), \(p=.012\)), consistent with prior findings on the beneficial correlates of flow.

\subsection{Social Media and Flow}
\subsubsection{Social Media Use Patterns}
\begin{figure*}[h]
  \centering
  \includegraphics[width=\linewidth]{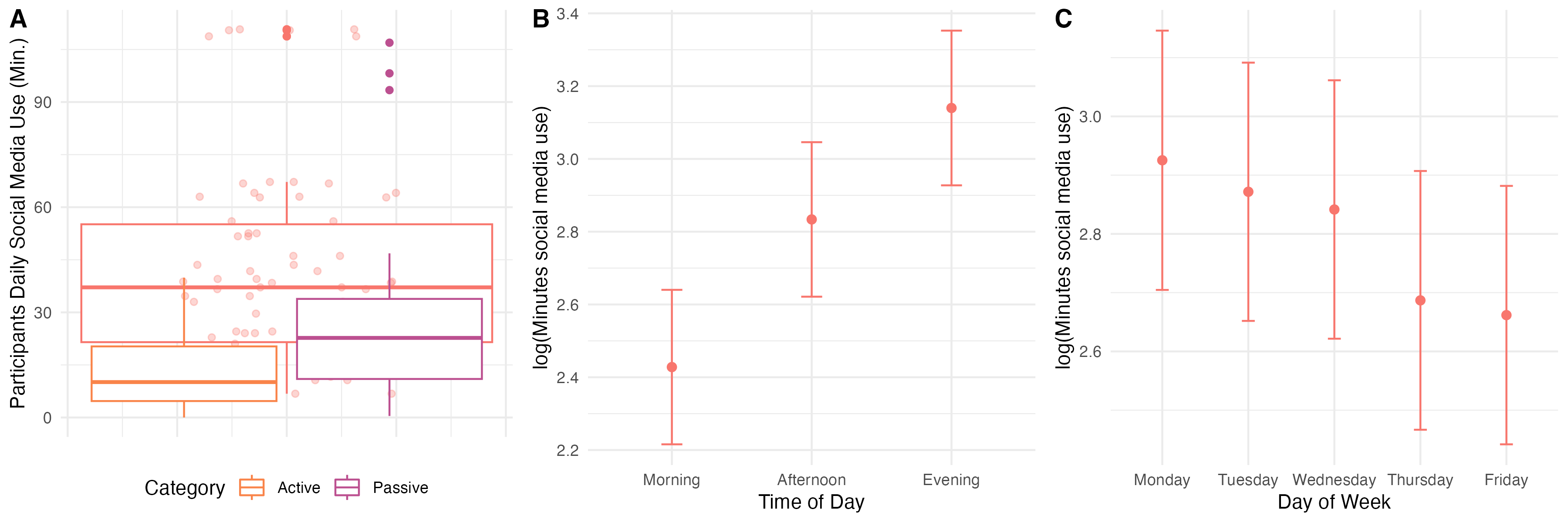}
  \caption{Patterns of social media use. Left: distribution of daily minutes in active (orange) and passive (pink) apps. Middle: average social media use across times of day. Right: average social media use across weekdays. Error bars indicate one SE.}
  \label{fig:SMUPatterns}
  \Description{The figure shows three panels of social media use data. The left panel compares active and passive use: active use takes fewer minutes per day with a narrower range, while passive use is more variable and sometimes much higher. The middle panel shows time-of-day patterns, with use lowest in the morning and increasing in the afternoon and evening. The right panel shows weekday patterns, with use roughly similar across days but slightly lower toward Friday.}
\end{figure*}

Social media was used an average of 41.93 minutes per day (13.64 minutes for active and 28.29 minutes for passive social media apps - see Figure \ref{fig:SMUPatterns}A). Variation of app use per participant was considerable (min avg. daily use: 6.78 minutes, max avg. daily use: 118.01 minutes).

Linear mixed models (with participant as random intercepts and log-transformed social media minutes to account for the skewed distribution) show that social media use significantly increases throughout the day (\(b=0.357\), \(SE=0.043\), \(p<.001\)) and a decrease throughout the week at trend level (\(b=-0.072\), \(SE=0.026\), \(p=.098\) - see Figure \ref{fig:SMUPatterns}B-C).

\begin{figure*}[h]
  \centering
  \includegraphics[width=0.45\linewidth]{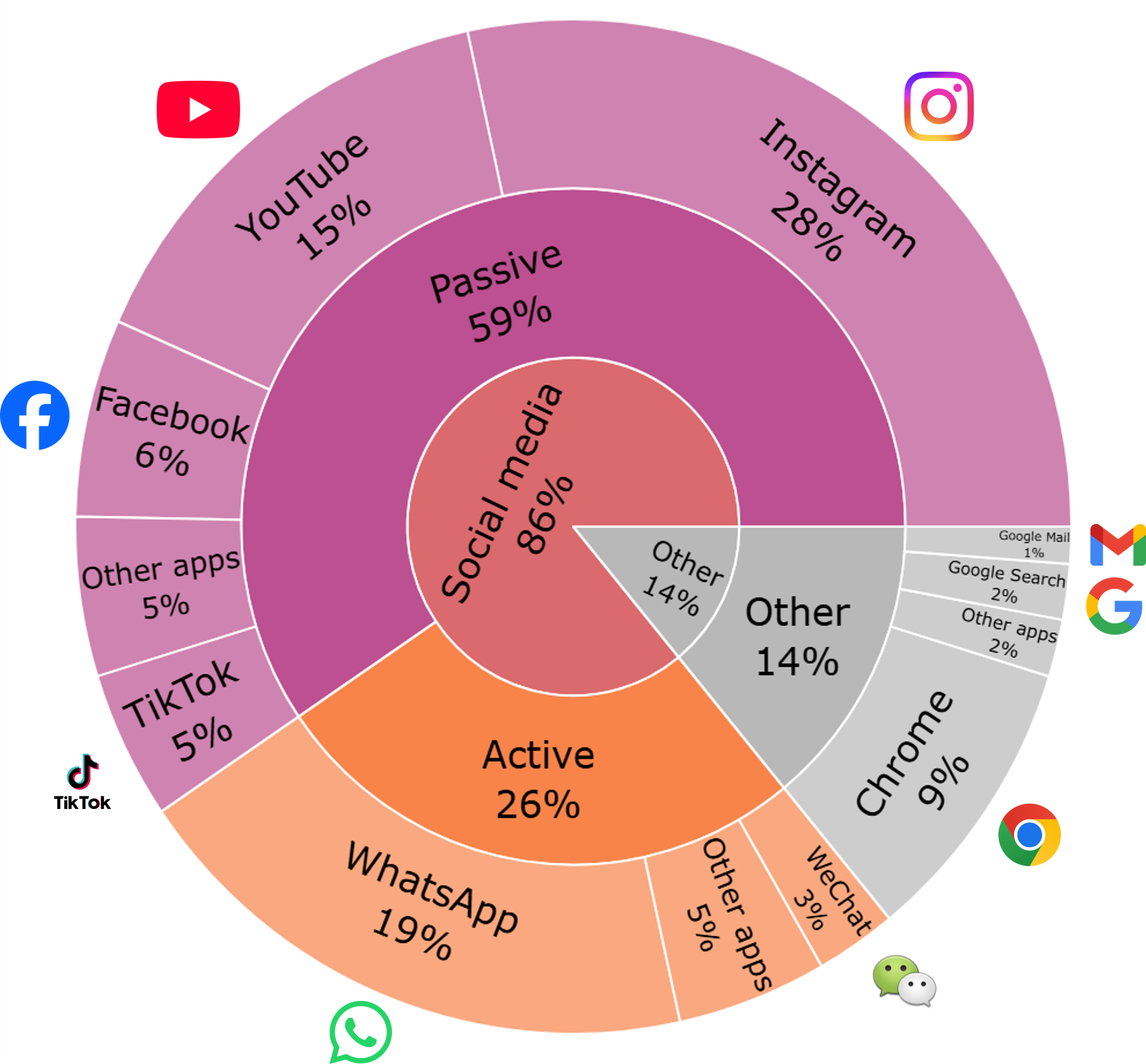}
  \caption{Distribution of app use across categories.}
  \label{fig:SMUDistributions}
  \Description{The figure is a sunburst chart showing how participants’ app use was distributed. Social media made up 86\% of all logged usage. Within this, passive apps (59\%) were the largest share, especially Instagram (28\%) and YouTube (15\%), with smaller contributions from Facebook and TikTok (5–6\% each). Active apps (26\%) were led by WhatsApp (19\%), followed by WeChat (3\%) and others. The remaining 14\% of use came from non-social apps, mostly Chrome (9\%) and Google services.}
\end{figure*}

Social media use in total accounted for 86\% of the tracked and used apps. Active social media accounted for 26\% of use, with the most frequent apps being WhatsApp (19\%), and WeChat (3\%). Passive social media accounted for 59\%, with Instagram (28\%), YouTube (15\%), and Facebook (6\%) being the most heavily used apps (more details in Figure \ref{fig:SMUDistributions}).

\subsubsection{Relationship with Flow}
\begin{figure*}[h]
  \centering
  \includegraphics[width=\linewidth]{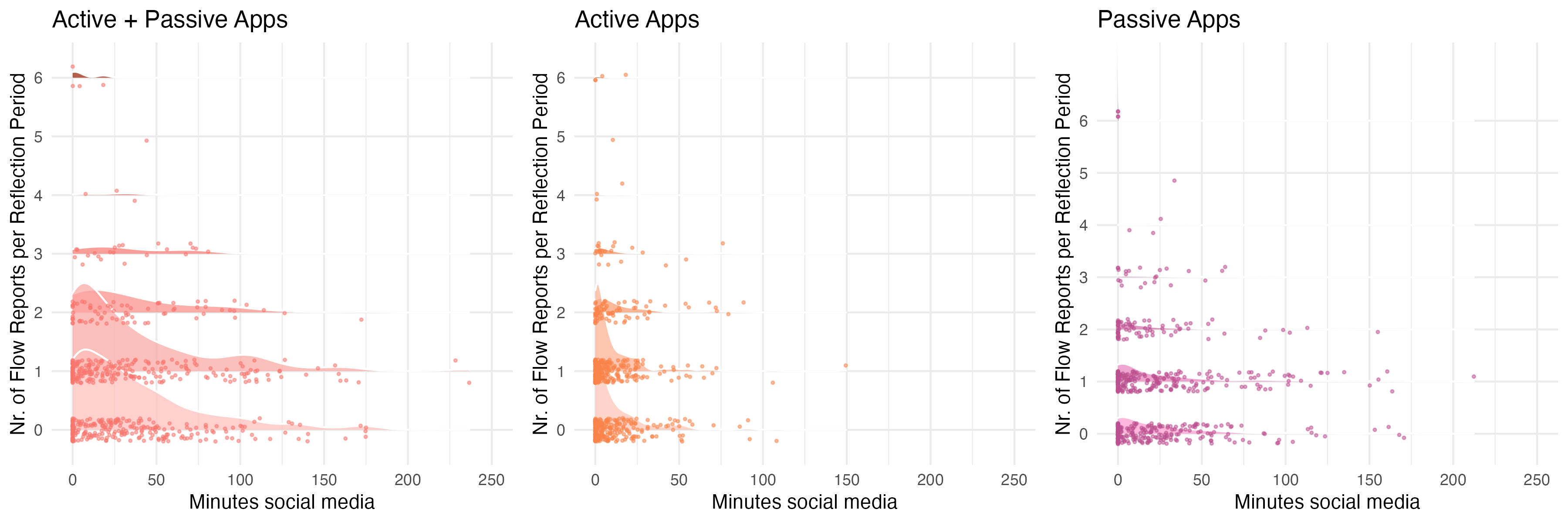}
  \caption{Relationship between social media use and flow reports. Each panel shows ordinal flow reports (y-axis) plotted against minutes of use (x-axis), with separate plots for total use (left), active use (middle), and passive use (right). Scatter points represent individual reflection sessions.}
  \label{fig:flowSMU}
  \Description{The figure has three panels, each plotting social media use in minutes on the horizontal axis against flow reports on the vertical axis. The left panel shows total use, the middle active use, and the right passive use. Across panels, most flow reports cluster at lower usage levels, especially below 50 minutes per reflection period, with fewer flow reports at higher levels of use. Passive use shows a particularly strong pattern, with higher use linked to fewer flow reports, while active use shows little systematic relationship.}
\end{figure*}

To analyze the relationship of social media use (duration in minutes) with flow occurrences (reported instances) - see Figure \ref{fig:flowSMU}, ordinal mixed regression models were used with random intercepts for participants and predictors separated for within- and between-person differences to assess whether potential differences in flow may arise from between-person differences or from daily behaviors (within-person differences). 
We analyzed total social media use and active and passive apps separately, as previous work highlighted that active social media use (e.g. through messaging with others) may have different cognitive-affective outcomes than passive social media use (e.g. scrolling and watching) \cite{Verduyn2015,Verduyn2017,Godard2024,Liang2024}.
In model development, we tested both time of day and day of week as covariates, but due to collinearity with time of day, only day of week was retained. The models are summarized in Table~\ref{tab:flowbysmu}. 

For overall social media use, higher within-person use predicted fewer flow reports (\(p=.003\)), with no reliable between-person effect (\(p=.083\)). Active apps showed no significant associations (\(p>.1\)). For passive apps, greater within-person use again predicted fewer flow reports (\(p=.002\)), with no between-person effect (\(p>.1\)).
Thus, altogether, people reported fewer flow experiences when they used social media more (during the day) - especially with social media apps that are categorized here as primarily passive.

\begin{table*}[h]
\caption{Cumulative logit mixed models for flow predicted by social media use duration.}
\label{tab:flowbysmu}
\centering
\begin{tabular}{lllllll}
\toprule
                           & \multicolumn{2}{l}{\textbf{\(\text{Flow} \sim   \text{All SM Apps}\)}} & \multicolumn{2}{l}{\textbf{\(\text{Flow} \sim   \text{Active Apps}\)}} & \multicolumn{2}{l}{\textbf{\(\text{Flow} \sim   \text{Passive Apps}\)}} \\
\midrule
\multicolumn{7}{l}{\textit{\textbf{Fixed Effects}}}                                                                                                                                                                                              \\
Minutes SMU Within-Person  & -0.531 (0.177)                     & p=.003**                      & -0.375 (0.346)                       & p=.277                         & -0.683 (0.224)                       & p=.002**                        \\
Minutes SMU Between-Person & 0.798 (0.454)                      & p=.083                        & 1.058 (1.147)                        & p=.356                         & 0.892 (0.523)                        & p=.088                          \\
Day of Week                & -0.396 (0.061)                     & p<.001***                     & -0.378 (0.060)                       & p<.001***                      & -0.393 (0.060)                       & p<.001***                       \\
\midrule
\multicolumn{7}{l}{\textit{\textbf{Random Effects}}}                                                                                                                                                                                             \\
Participant (Lvl-1)        & \multicolumn{2}{l}{1.365***}                                       & \multicolumn{2}{l}{1.339***}                                          & \multicolumn{2}{l}{1.372***}                                           \\
\midrule
\multicolumn{7}{l}{\scriptsize{\textit{\textbf{Note:}   Random effects shown as variance; $^{***}p<.001$; $^{**}p<.01$;   $^{*}p<.05$}}} \\
\bottomrule         
\end{tabular}
\end{table*}

\subsubsection{Pathways for the Influence of SMU on Flow}
To explore mechanisms underlying the negative within-person link between social media use and flow, we examined two candidate factors: mental fatigue and mental readiness (a composite of mood, motivation, and sleepiness). Internal consistencies for both variables were high (Cronbach’s Alpha = .92 for fatigue, .78 for readiness) and they showed a strong negative correlation (\(r(553)=-0.69\), \(p<.001\)). This is in line with previous research that confirms this correlation, yet also highlights a distinct factor structure \cite{Schick2025}.

Given their correlation, we modeled both factors in separate ordinal mixed-effects regressions (Table~\ref{tab:flowbyfatigueandreadiness}), decomposing predictors into within-person (session-level deviations from each participant’s mean) and between-person (participant means across days).  
Results showed that higher fatigue was weakly associated with fewer flow reports (\(p=.059\)) at the within-person level, with no between-person effect (\(p>.1\)). In contrast, higher readiness was strongly associated with more flow (\(p<.001\)) within persons, again with no between-person effect (\(p>.1\)).

\begin{table*}[h]
\caption{Cumulative logit mixed models for flow predicted by mental fatigue and readiness.}
\label{tab:flowbyfatigueandreadiness}
\centering
\begin{tabular}{lllll}
\toprule
                     & \multicolumn{2}{l}{\textbf{\(\text{Flow} \sim   \text{Mental Fatigue}\)}} & \multicolumn{2}{l}{\textbf{\(\text{Flow} \sim   \text{Mental Readiness}\)}} \\
\midrule
\multicolumn{5}{l}{\textit{\textbf{Fixed Effects}}}                                                                                                                            \\
State Within-Person  & -1.000 (0.503)                         & p=.059                           & 2.478 (0.657)                           & p<.001***                         \\
State Between-Person & -0.153 (1.472)                         & p=.916                           & 1.928 (1.589)                           & p=.225                            \\
Day of Week          & -0.375 (0.059)                         & p<.001***                        & -0.374 (0.060)                          & p<.001***                         \\
\midrule
\multicolumn{5}{l}{\textit{\textbf{Random Effects}}}                                                                                                                           \\
Participant (Lvl-1)  & \multicolumn{2}{l}{1.362***}                                              & \multicolumn{2}{l}{1.362***}                                                \\
\midrule
\multicolumn{5}{l}{\scriptsize{\textit{\textbf{Note:}   Random effects shown as variance; $^{***}p<.001$; $^{**}p<.01$;   $^{*}p<.05$}}}                                      \\
\bottomrule
\end{tabular}
\end{table*}

We next modeled fatigue (Table~\ref{tab:fatiguebysmu}) and readiness (Table~\ref{tab:readinessbysmu}) as outcomes of social media use, again separating within- and between-person effects and social media app categories.  
For fatigue, greater within-person use was associated with higher fatigue (\(p=.030\)), while between-person use was not significant. When disaggregated by type, the within-person effect for active use \revision{was not significant} (\(p=.080\)), and no effects were found for passive use (\(p>.1\)).  

For readiness, greater within-person use predicted lower readiness (\(p=.004\)), whereas higher between-person use predicted higher readiness (\(p=.009\)). By app category, active apps showed only the negative within-person association (\(p=.013\)), while passive apps showed both the negative within-person (\(p=.041\)) and the positive between-person (\(p=.024\)) associations.  

\begin{table*}[h]
\caption{Linear mixed models for mental fatigue predicted by social media use.}
\label{tab:fatiguebysmu}
\centering
\begin{tabular}{lllllll}
\toprule
                           & \multicolumn{2}{l}{\textbf{\(\text{Fatigue} \sim   \text{All SM Apps}\)}} & \multicolumn{2}{l}{\textbf{\(\text{Fatigue} \sim   \text{Active Apps}\)}} & \multicolumn{2}{l}{\textbf{\(\text{Fatigue} \sim   \text{Passive Apps}\)}} \\
\midrule
\multicolumn{7}{l}{\textit{\textbf{Fixed Effects}}}                                                                                                                                                                                                       \\
Minutes SMU Within-Person  & 0.033 (0.015)                         & p=.030*                       & 0.052 (0.029)                           & p=.080                         & 0.031 (0.019)                           & p=.109                          \\
Minutes SMU Between-Person & -0.068 (0.048)                        & p=.168                        & -0.088 (0.125)                          & p=.483                         & -0.068 (0.055)                          & p=.222                          \\
Day of Week                & 0.003 (0.005)                         & p=.950                        & -0.001 (0.050)                          & p=.978                         & -0.001 (0.050)                          & p=.992                          \\
\midrule
\multicolumn{7}{l}{\textit{\textbf{Random Effects}}}                                                                                                                                                                                                      \\
Participant (Lvl-1)        & \multicolumn{2}{l}{0.017***}                                          & \multicolumn{2}{l}{0.018***}                                             & \multicolumn{2}{l}{0.018***}                                              \\
\midrule
\multicolumn{7}{l}{\scriptsize{\textit{\textbf{Note:}   Random effects shown as variance; $^{***}p<.001$; $^{**}p<.01$;   $^{*}p<.05$}}} \\
\bottomrule
\end{tabular}
\end{table*}

\begin{table*}[h]
\caption{Linear mixed models for mental readiness predicted by social media use.}
\label{tab:readinessbysmu}
\centering
\begin{tabular}{lllllll}
\toprule
                           & \multicolumn{2}{l}{\textbf{\(\text{Readiness}   \sim \text{All SM Apps}\)}} & \multicolumn{2}{l}{\textbf{\(\text{Readiness}   \sim \text{Active Apps}\)}} & \multicolumn{2}{l}{\textbf{\(\text{Readiness}   \sim \text{Passive Apps}\)}} \\
\midrule
\multicolumn{7}{l}{\textit{\textbf{Fixed Effects}}}                                                                                                                                                                                                             \\
Minutes SMU Within  & -0.035 (0.012)                        & p=.004**                        & -0.058 (0.023)                          & p=.013*                          & -0.030 (0.015)                           & p=.041*                          \\
Minutes SMU Between & 0.117 (0.042)                         & p=.009**                        & 0.174 (0.114)                           & p=.134                           & 0.115 (0.049)                            & p=.024*                          \\
Day of Week                & -0.004 (0.004)                        & p=.255                          & -0.004 (0.004)                          & p=.305                           & -0.004 (0.004)                           & p=.302                           \\
\midrule
\multicolumn{7}{l}{\textit{\textbf{Random Effects}}}                                                                                                                                                                                                            \\
Participant (Lvl-1)        & \multicolumn{2}{l}{0.014***}                                            & \multicolumn{2}{l}{1.339***}                                               & \multicolumn{2}{l}{1.372***}                                                \\
\midrule
\multicolumn{7}{l}{\scriptsize{\textit{\textbf{Note:}   Random effects shown as variance; $^{***}p<.001$; $^{**}p<.01$;   $^{*}p<.05$}}}  \\
\bottomrule
\end{tabular}
\end{table*}

\begin{figure*}[h]
  \centering
  \includegraphics[width=0.85\linewidth]{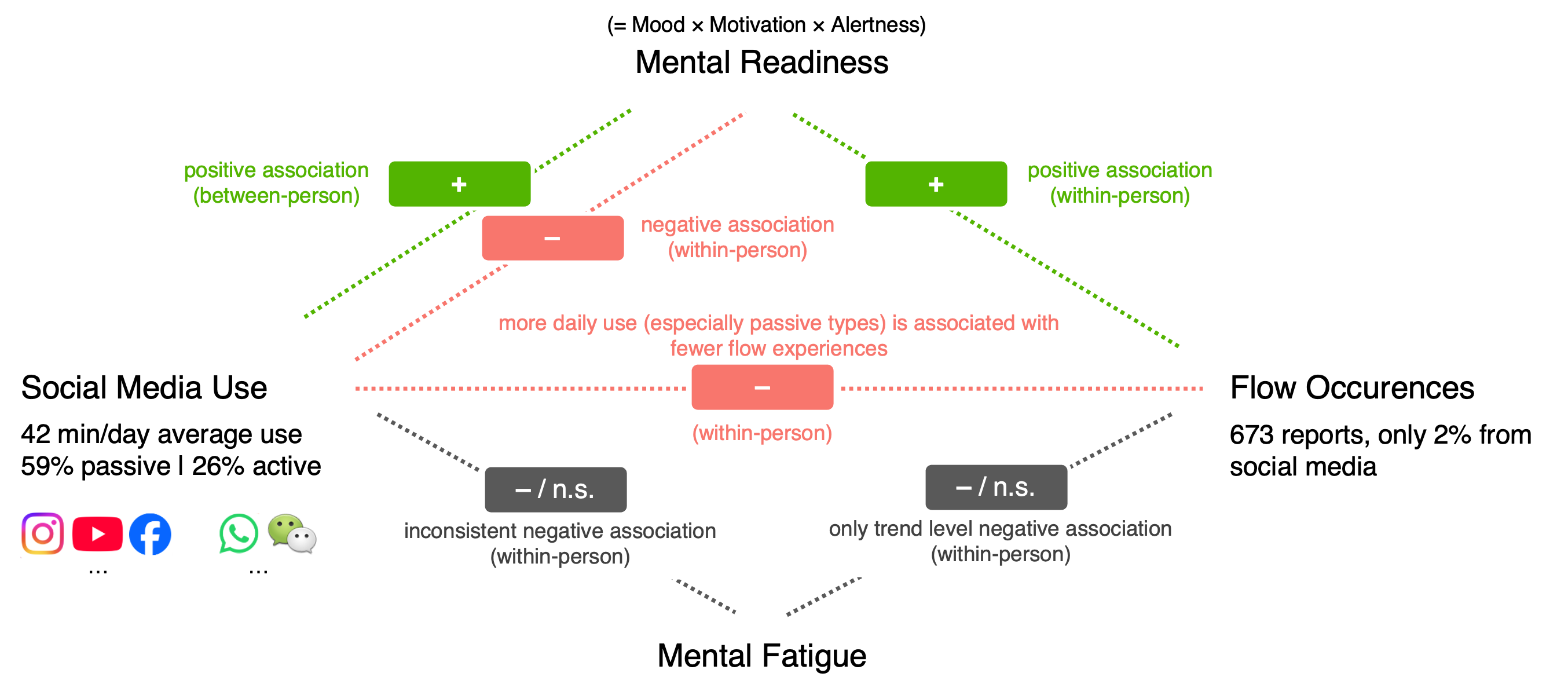}
  \caption{\revisionfinal{Results Summary: Dashed lines indicate statistical associations (non-causal). Green denotes positive associations, red negative associations, and grey weak or non-significant associations. Between-person effects were not significant unless stated.
  Flow occurred rarely during social media use. Greater daily social media use (particularly passive app types) was associated with fewer flow experiences within individuals.
  Greater daily social media use was negatively associated with mental readiness, and higher daily mental readiness was related to more frequent flow occurrences. Higher habitual social media use was associated with generally higher mental readiness.
  Associations between social media use and mental fatigue were inconsistent and weak, and mental fatigue showed only a trend-level negative association with flow.}}
  \label{fig:resultsOverview}
  \Description{The figure is a conceptual diagram summarizing associations between four constructs: Social Media Use (left), Mental Readiness (top center), Mental Fatigue (bottom center), and Flow Occurrences (right). Mental readiness is defined as the combination of mood, motivation, and alertness.
  Dashed arrows indicate statistical associations (not causal). Green arrows represent positive associations, red arrows negative associations, and grey arrows weak or non-significant associations.
  Within individuals, higher daily social media use—especially passive use—is negatively associated with mental readiness and with flow occurrences. Mental readiness is positively associated with flow within individuals. Between individuals, higher habitual social media use is positively associated with higher mental readiness.
  Associations between social media use and mental fatigue are weak and inconsistent within individuals. Mental fatigue shows only a weak, trend-level negative association with flow.
  Additional text indicates that participants used social media on average 42 minutes per day, with 59\% passive use and 26\% active use, and that flow was rarely reported during social media activities (about 2\% of all 673 flow reports). Social media platforms (e.g., Instagram, YouTube, Facebook, WhatsApp, WeChat) are shown as examples.}
\end{figure*}

Overall, associations between social media, flow and fatigue were weak and inconsistent, whereas readiness showed a clearer cross-level pattern. Within persons, using more social media than one’s own typical level was linked to lower readiness, and higher readiness coincided with more flow. Between persons, heavier typical use was associated with higher average readiness. This sign flip suggests selection or role/routine effects: people who generally use social media more may also have traits or contexts that keep them readier on average, even though for any given individual, using more than usual aligns with feeling less ready. The pattern was strongest for passive apps (negative within-person, positive between-person), while for active apps only the negative within-person effect remained. Altogether, readiness - not fatigue - emerges as the more plausible pathway between social media use and flow, though the evidence remains correlational at this point. \revision{Our key results are summarized in Figure \ref{fig:resultsOverview}.}

\section{Discussion}
Our findings challenge the prevalent assumption that social media serves as a regular source of flow in daily life. Two key results warrant critical discussion: the rarity of flow attribution to social media use, and the negative within-person relationship between social media engagement and flow occurrence.

\subsection{Flow Attributions to Social Media}
Participants spontaneously identified social media as flow-inducing in only \(\sim\)2\% of reports (14/673), contrasting with claims that platforms are (naturally or purposefully) designed to facilitate flow-like engagement \cite{Zhou2025,Chen2025}. When reflecting freely, flow was overwhelmingly attributed to other pursuits - academic work, leisure, physical activity, and social interactions offline or via active communication rather than passive consumption. 

We find this low prevalence credible for several reasons. First, the DRM captured unprompted recollections across multiple daily windows, reducing the chance that social-media flow was systematically missed. Second, the large number of flow reports indicates participants engaged with the task; they simply did not link flow to social media. Third, objective logs confirmed substantial use (avg. 42 minutes/day), ruling out minimal exposure as an explanation. Lastly, these results replicate earlier flow research that finds passive activities like watching TV rarely elicit vivid flow \cite{Csikszentmihalyi1989, Engeser2016}.

However, alternative accounts deserve consideration: social desirability may discourage attributing flow to stigmatized or “excessive” screen behaviors, a phenomenon previously observed in the social media use literature \cite{Boyle2022,Coyne2023}. Moreover, memory limitations could make fast, fragmented flow sessions less salient. Yet these explanations struggle to account for the magnitude of the effect: leisure like gaming and entertainment appeared more often in flow reports, and social media frequently features emotionally engaging content that should be memorable.

In our view, a more plausible interpretation is that the low attribution reflects genuine phenomenological differences between social media engagement and flow states. Classic flow theory emphasizes clear goals, immediate feedback, and challenge-skill balance \cite{Csikszentmihalyi1975}. While social media platforms provide immediate feedback through likes and comments, the goals are often ambiguous (what constitutes "enough" scrolling?) and the challenge minimal (consumption requires little skill development). The resulting experience may produce a "pseudo-flow": absorption and time distortion - experiences superficially similar to flow - yet without the goal-directed engagement and skill mastery that characterize authentic flow states \cite{Abuhamdeh2020}. 
\revision{Empirically separating pseudo-flow from genuine flow may be difficult, but multi-item flow measures coupled with indicators of challenge and active engagement could enable this. Relevant indicators could include behavioral patterns that differentiate goal-directed activity from passive consumption, and physiological signals reflecting attentional dynamics (e.g., EEG \cite{Chiossi2025,Knierim2025}).
Recognizing this distinction also opens design opportunities grounded in recent intervention studies. Instead of maximizing passive consumption, platforms could use indications of pseudo-flow to steer extended browsing toward more intentional actions by foregrounding social or creative actions such as picking up an ongoing conversation, resuming a project, or curating content \cite{Landesman2024,Skeggs2025,Purohit2023}. Distinguishing pseudo-flow from flow thus clarifies engagement dynamics and suggests ways to channel usage into more skill-involving, meaningful, and connection-oriented interactions that benefit both users and providers.}

\subsection{Social Media Use as Flow Competitor Rather Than Facilitator}
Besides rare occurrence, our analyses also indicate a negative link between within-day social media use and flow. 
Two plausible pathways could underlie the negative association between social media use and flow: fatigue and readiness (especially through mood and motivation). Periods with more app use than typical were linked to lower readiness, and lower readiness in turn coincided with fewer flow reports - suggesting that social media may undermine conditions for flow rather than merely failing to provide them. This interpretation aligns with experimental evidence that brief social media sessions can impair vigilance/inhibition and increase subjective fatigue \cite{Gantois2021,Jacquet2023}, and with design features - rapid context switching and social comparison - that fragment attention and depress affect \cite{Verduyn2015,Verduyn2017}. Even when use feels engaging, it may leave users less able to achieve flow in subsequent activities. Yet, we also do not claim mediation - our observed effects are correlational and concurrent, but the results outline testable pathways for future work (e.g., time-lagged models).

Also, alternative explanations remain plausible. One is reverse causation: people may turn to social media when they already feel low in readiness or unable to enter flow, using it as a default when more demanding activities feel overwhelming \cite{Oulasvirta2012,Reinecke2009}. Our design helps to address this concern, as social media use and flow were reconstructed for the preceding hours, while readiness and fatigue were measured at the time of the reflection. This ordering reduces the risk that low readiness at the moment of reporting simply biases which activities are recalled as flow-inducing. Still, we cannot fully establish temporal precedence, and future work should strengthen causal claims by applying lagged analyses and higher-frequency sampling.
A second explanation is displacement: social media may reduce flow not via direct psychological effects, but by substituting for activities that are more likely to elicit it. Longer scrolling periods often arise in precisely those discretionary windows when active leisure - pursuing hobbies, creative practice, or skill-based recreation - could provide clear goals, feedback, and challenge–skill balance \cite{Csikszentmihalyi1989,Engeser2016}. In this case, the opportunity cost is high: passive media consumption might fill opportunities that might otherwise be invested in flow-conducive pursuits, thereby lowering the day’s overall chances to experience flow.

\revision{Our findings connect to broader debates in digital health, wellbeing and related technology ecologies \cite{Valkenburg2022,Roffarello2023}. These literatures emphasize that outcomes depend not on use durations, but on whether technologies afford eudaimonic (meaning-making, growth) versus hedonic (pleasure-seeking) experiences \cite{Peters2018, Valkenburg2022} and how digital tools fit into broader activity systems \cite{Epstein2015,Rooksby2014,Roffarello2023}. 
Our results suggest that everyday social media use rarely affords eudaimonic states like flow and may, at higher intensities, compete with more flow-conducive activities. At the same time, social media can support restorative functions \cite{Kim2019}, and people need rest to experience flow \cite{Debus2014}. Thus, our contribution lies less in documenting harm than in highlighting the scarcity of eudaimonic experiences within social media use, consistent with recent work finding social media's role in wellbeing ambiguous \cite{Valkenburg2022}. Within daily activity ecologies, flow appears tied to more active pursuits – learning, creating, conversing – which often require sustained engagement rather than brief episodes \cite{Csikszentmihalyi1989}. This temporal dimension matters: short social media breaks are unlikely to displace flow, whereas extended scrolling during prime discretionary periods may do so. Our key question for digital wellbeing research is, therefore, not how much social media people use, but whether and how platforms can support more eudaimonic, skill-involving experiences. Emerging generative AI tools could play a role here by lowering barriers to creative production and serving as "creator's companions" for challenging projects - though realizing this potential requires design that supports genuine skill-building rather than passive content generation ("AI slop").}

\subsection{Limitations and Future Directions}
Our findings should be interpreted in light of several methodological and contextual constraints, including the scope of device logging, the retrospective nature of the DRM, and the granularity of our social media measures.

\paragraph{Device coverage and rationale.}
Our telemetry captured social media use on participants’ smartphones, but not on tablets, computers, or TVs. Smartphones sit at the center of multi-device ecologies, anchoring short, frequent social media checks \cite{Heitmayer2022}. Smartphones are also the primary setting for habitual “mobile social media” use linked to task-unrelated thoughts and attention costs \cite{Meier2023}, and, in constrained contexts like school/work hours, objectively logged phone activity is dominated by social apps \cite{Burnell2025}. Focusing on phones further enabled time-locked linkage to experience sampling, and recent work highlights the value of device logs over self-reports \cite{Coyne2023,Sewall2022}. Our estimates should thus be read as the \emph{smartphone slice} of social media use - well-suited to capture ubiquitous, short-burst interactions most likely to affect within-day flow and readiness, but undercounting creation-oriented or lean-back sessions more common on larger screens. Future work could extend this with cross-device logging.

\paragraph{DRM method limitations.}
The DRM decouples experience from attribution but emphasizes memorable, emotionally salient episodes, potentially underrepresenting brief or subtle states \cite{Lucas2021}. This bias is less problematic for flow - long described as a cognitively salient experience \cite{Csikszentmihalyi1975} - though experience sampling (ESM) has more often been used in flow research \cite{Csikszentmihalyi2003,Csikszentmihalyi1989}. ESM may capture fleeting flow during scrolling or micro-interactions, but its random prompts reduce daily coverage \cite{Bartholomeyczik2024}. Overall, the DRM’s focus on memorable episodes fits our theoretical interest in flow while avoiding attribution biases common in social media research. 
\revision{Second, both DRM and ESM involve repeated reflection on one’s experiences, which may introduce some reactivity - heightening awareness or subtly adjusting behavior over time \cite{Ottenstein2024}. Such effects could extend indirectly to related behaviors (e.g., social media use). To that end,} future work could substitute DRM and ESM through multimodal classifiers (e.g., HRV, EDA, EEG) linked to app telemetry to provide continuous, in-situ indicators of flow, though this will require careful ground-truthing and validation \cite{Rissler2020,Knierim2025}.

\revision{\paragraph{Study design and operationalization.}
Our correlational design limits causal interpretations. While higher social media use coincided with fewer flow reports, we cannot establish whether use might even directly reduce flow, whether lower readiness drives both, or whether contextual factors explain the pattern. 
Future work could adapt the designs of prior social-media‐use interventions by introducing randomized time-limited \cite{Olson2023} use conditions in everyday life and measuring subsequent flow experiences - although previous results suggest that such interventions will require more than one week of use \cite{Wezel2021}.
Furthermore, our sample size (N=44) lies toward the lower end of ranges used in prior flow sampling studies (~20–120 participants - \cite{Debus2014, Bartholomeyczik2024, Brown2023}), and a larger cohort would improve generalizability, as would an even more diverse sample. At the same time, the core patterns are showing large effects - for example, only $\sim$2\% of all recorded flow episodes were attributed to social media - supporting the credibility of our main conclusions despite the sample size limitation.}

\paragraph{Social media context and variety.}
Our aggregate treatment of social media obscures differences between behaviors within apps, specific features, and contexts. Whereas the active–passive distinction is typically defined at the behavior level \cite{Burke2011,Verduyn2017}, we used app categories as a pragmatic proxy, which inevitably mixes behaviors and should be interpreted cautiously \cite{Schoedel2022}. In our data, minutes in feed/video apps were more consistently associated with fewer flow reports than minutes in messaging apps, suggesting that a broad separation can be informative - but coarse and potentially attenuated by misclassification.

Flow potential likely also varies within platforms - e.g., collaborative creation or reciprocal exchanges may support flow, while algorithmic feeds may undermine it - and depends on user motives such as information seeking, entertainment, or social connection \cite{Whiting2013}. Our aggregation was necessary to establish baseline prevalence without cherry-picking platforms or features, showing that across typical social media engagement, spontaneous flow reports were rare. Future work should refine this baseline by combining app telemetry with within-app action traces or behavior items \cite{Gerson2017,Karsay2023} to identify which features and contexts, if any, reliably support flow.

\section{Conclusion}
We examined whether social media meaningfully contributes to everyday flow by pairing app-level telemetry with unprompted flow reports over five days. Our findings reveal a clear disconnect: Participants rarely identified social media as flow-inducing (2\% of reports), and higher-than-usual use predicted fewer flow occurrences within individuals.  
These results challenge the assumption that social media reliably elicits flow. When reflecting freely, participants attributed flow mainly to academic work, creative pursuits, physical activity, and face-to-face interactions. The negative within-person link suggests that social media often competes with, rather than facilitates, optimal experience - whether through direct psychological effects or by displacing flow-conducive activities.  

This pattern does not imply that all social media use is inherently problematic for flow. People may benefit from breaks between demanding, flow-inducing activities, and platforms clearly serve other valued functions including social connection, entertainment, and information access. Yet, the critical point we want to make is that engagement should not be conflated with optimal experience.

For HCI research, our findings underscore the methodological value of approaches that decouple experience from attribution. For platform design, they suggest prioritizing metrics that capture experience quality rather than time-on-platform alone. Future research should continue probing when, for whom, and under what specific conditions social media might genuinely support flow - while recognizing that its general relationship to optimal experience appears weak and potentially even negative. Understanding these nuances becomes increasingly important as digital platforms play ever-larger roles in shaping daily human experience.

\begin{acks}
Funded by the German Research Foundation (Deutsche Forschungsgemeinschaft - DFG) – GRK2739/1 – Project Nr. 447089431 – Research Training Group: KD²School – Designing Adaptive Systems for Economic Decisions.
\end{acks}

\bibliographystyle{ACM-Reference-Format}
\bibliography{main}


\end{document}